# Seeing virtual while acting real: visual display and strategy effects on the time and precision of eye-hand coordination

Anil U. Batmaz[1], Michel de Mathelin[1], Birgitta Dresp-Langley[1]

[1]ICube Laboratory, University of Strasbourg, CNRS
1, Place de l'Hopital
67000 Strasbourg, FRANCE

#Corresponding author; E-mail: birgitta.dresp@unistra.fr

This work was financed by the IDEX (Initiative D'EXcellence) of Strasbourg University with a grant allowance to B. Dresp-Langley for her project on "Tactile exploration of complex object properties with and without direct visual guidance". A. U. Batmaz was hired as doctoral student on that project within the ICube Lab UMR 7357 CNRS and University of Strasbourg. The experimental platform *"EXCALIBUR"* was financed by a grant to B. Dresp (AAP MI CNRS 2015).

# Abstract

Effects of different visual displays on the time and precision of bare-handed or tool-mediated eye-hand coordination were investigated in a pick-and-place-task with complete novices. All of them scored well above average in spatial perspective taking ability and performed the task with their dominant hand. Two groups of novices, four men and four women in each group, had to place a small object in a precise order on the centre of five targets on a Real-world Action Field (RAF), as swiftly as possible and as precisely as possible, using a tool or not (control). Each individual session consisted of four visual display conditions. The order of conditions was counterbalanced between individuals and sessions. Subjects looked at what their hands were doing 1) directly in front of them ("natural" top-down view) 2) in top-down 2D fisheye view 3) in top-down undistorted 2D view or 4) in 3D stereoscopic top-down view (head-mounted OCULUS DK 2). It was made sure that object movements in all image conditions matched the real-world movements in time and space. One group was looking at the 2D images with the monitor positioned sideways (sub-optimal); the other group was looking at the monitor placed straight ahead of them (near-optimal). All image viewing conditions had significantly detrimental effects on time (seconds) and precision (pixels) of task execution when compared with "natural" direct viewing. More importantly, we find significant trade-offs between time and precision between and within groups, and significant interactions between viewing conditions and manipulation conditions. The results shed new light on controversial findings relative to visual display effects on eye-hand coordination, and lead to conclude that differences in camera systems and adaptive strategies of novices are likely to explain these.

# Introduction

In image-guided processes for decision and action, as in laparoscopic surgical interventions, the human operator has to process critical information about what his/her hands are doing in a real-world environment while looking at a two-dimensional (2D) or three-dimensional (3D) representation of that environment displayed on a monitor. This virtual information needs to be correctly interpreted by the brain to ensure safe and effective human intervention [1, 2, 3, 4]. In comparison with direct observation and action, image-guided eye-hand coordination represents a disadvantage [5, 6], for essentially three reasons. First, veridical information about real-world depth is missing from the image representations. Second, the operator is looking sideways or straight ahead at a monitor, or at an image displayed by a head-mounted device, instead of looking down on his/her hands. Third, due to a variety of camera and image calibration problems, the hand or tool movements displayed virtually may not match the real-world movements in time and space.

The loss of higher order (cortical) depth cues in image-guided manual tasks has been identified as a major drawback, significantly affecting performances of novices compared with task execution in direct binocular or monocular vision [6, 7, 8]. Adapting to this constraint is possible through a long period of training to optimize indirect eye-hand coordination [2, 3], and surgeons who have this kind of training complete image-guided tasks significantly faster than novices, with significantly fewer tool movements, shorter tool paths, and fewer grasp attempts [9]. Developing this expertise requires significant adjustments in individual goal-control strategies [8, 9, 10]. While surgeons experienced in image-guided interventions tend to focus their attention on target locations, novices split their attention between trying to focus on targets and, at the same time, trying to track the tools [9]. This reflects a common strategy for controlling goal-directed hand movements in non-trained subjects and affects task execution times [11].

Recently developed 3D visualization technology may represent a possibility for overcoming the drawbacks of 2D views, yet, whether different 3D imaging solutions all significantly improve task performance has remained a controversial issue. While some authors have reported that 3D viewing significantly improves task performance in both novices and experts [12-18], others have found equivalent task performance comparing 2D viewing to 3D viewing systems [19-22]. It has been suggested that differences in task complexity and inherent affordance levels [10, 23], or inter-individual differences in adaptive goal-setting strategies of novices [8] may account for differences in results between studies using similar 3D viewing systems.

The most recent results available in the dedicated literature come from the study by Sakata and colleagues. These authors [18] used a laparoscopic (Olympus Endoeye Flex) HD camera system that can be switched from 2D to 3D stereoscopic viewing mode. This system gets rid of problems relative to viewing position and viewing distance [24], and it is reported that under such conditions, the 3D viewing mode produces better depth judgments and faster task execution in both novice and expert surgeons.

Monitor position [25] matters in as far as a considerable misalignment of the eye-hand-target axis during task execution, caused by a sub-optimal monitor position constraining the operator to turn his/her head sideways during an intervention, significantly affects measures of postural comfort [26, 27], and interventional safety [28]. A monitor placed straight ahead of the operator, in line with the forearm-instrument motor axis and at a height lower than the eye-level when looking straight ahead, is recognized as the recommended optimal standard [25].

Spatial and/or temporal mismatches between images and real-world data may occur in monitor views generated by different camera types. Surgical fisheye lens cameras, for example, provide a hemispherical focus of vision with poor off-axis resolution and aberrant shape contrast effects at the edges of the objects viewed [29]. Current prototype research struggles to find camera solutions which provide a larger, corrected focus of vision compared with that of commonly used laparoscopic cameras [30]. Whether fisheye image views affect eye-hand coordination performance to a greater extent than undistorted 2D views is not known. Furthermore, spatial as well as temporal mismatches between movements viewed on the monitor and the corresponding real-world movements may occur as a consequence of specific constraints for placing the camera. In the case of endoscopic surgery, for example, the camera moves with the tool used to perform the intervention, and movements represented visually are not aligned with the surgeon's real arm and hand movements. Another problem with camera-monitor systems for technology-driven visuo-motor tasks consists of temporal asynchronies between frames of reference for vision and action [31, 32]. These are known to produce a cognitive phenomenon called visual-proprioceptive mismatch, which negatively affects task performance [33]. Cognitive mismatch of relative distances in virtual reality representations of large-scale environments to their real-world counterparts produce wrong turns in navigation tasks [34]. To overcome this drawback, the operator needs to work out a way of compensating for the mismatched cues and, as a consequence, feels less in control of his/her actions [35, 36]. Experienced surgeons learn to cope with this problem through training, but the cognitive mechanisms of this adaptation are not understood.

When a tool is used to manipulate physical objects, as in laparoscopic interventions, there is no direct tactile feed-back from the object to the sensory receptors in the hand [37], which communicate with cortical neurons driven by multisensory input [38-40]. Visual-haptic mapping for cross-modal sensory integration [41] is affected by such lack of direct sensory feed-back. Dynamic changes in cognitive hand and body schema representations [42-46] occur as a consequence, and these cognitive changes are consolidated by repeated tool-use [47]. They form a part of the processes through which operators acquire specific motor skills and experience by learning [31] to adapt to the constraints of laparoscopic interventions in long periods of training. For lack of experience in this specific process of tool-mediated eye-hand coordination, the performances of novices can be expected to be slower and less precise in tool-mediated object manipulation compared with the "natural" situation where they are using their hands directly and their skin receptors are in touch with the manipulated object.

In this study here, a five-step pick-and-place-task [48] is executed by complete novices using their bare dominant hand or a tool. We selected individuals with no surgical experience at all scoring high in spatial 3D perspective taking ability [49] to eliminate, as much as possible, hidden sources of potentially relevant eye-hand coordination skill variations in surgical study populations. A large majority of previous findings in this field were obtained with populations of surgeons with or without laparoscopic training, divided into “novices” and “experts” on that basis. Some relevant hidden sources of skill variability may have been left unaccounted for given that all surgeons share expertise in surgical eye-hand coordination procedures. The homogeneity of such experience in a “novice” study group may be difficult if not impossible to control. In our novice population here, all individuals are absolute beginners in image-guided eye-hand coordination and, in addition, they have no other potentially relevant surgical eye-hand coordination expertise. Here, we compare the effects of 2D image viewing with near-optimally and sub-optimally positioned monitors to the effects of direct "natural" 3D and to the effects of stereoscopic 3D viewing through a head-mounted display. Effects on both the time and the precision of task execution are assessed. The head-mounted 3D system gets rid of problems relative to both viewing position and viewing distance.

On the basis of results from previous work [6-8], it is predicted that “natural” top-down direct viewing will produce the best task performance for time and precision compared with top-down 2D image views (fisheye or corrected). As predicted by other [21, 25, 28], a sub-optimal monitor position, where the subject has to look sideways to perform the task, is predicted to affect task performance negatively compared with a near-optimal viewing position, where the monitor is aligned with the fore-arm motor task execution axis and the subject is looking straight ahead. Our head-mounted 3D system presenting the same advantage of controlling for effects of viewing angle and distance as the 3D stereo system in Sakata et al [18], we expect faster task execution times compared with 2D views from a near-optimally placed monitor. In our system, the stereo view is generated by two HD fisheye cameras at fixed locations, while in the display used by Sakata et al [18], the endoscopic HD camera producing the images for left and right moves along with the tool.

# Materials and methods

We built a computer controlled perception-action platform (*EXCALIBUR*) for image-based analysis of data relative to the time and precision of real-world manual operations, performed by non-trained, healthy adult men and women volunteers under different conditions of object manipulation and 2D or 3D viewing.

*Ethics*

The study was conducted in conformity with the Helsinki Declaration relative to scientific experiments on human individuals with the full approval of the ethics board of the corresponding author's host institution (CNRS). All participants were volunteers and had provided written informed consent. The individual shown in Figure 1 of this manuscript has given written informed consent (as outlined in the PLOS consent form) to publish her picture.

*Subjects*

Eight healthy right-handed men ranging in age between 25 and 45, and eight healthy right-handed women ranging in age between 25 and 45 participated in this study. They were all highly achieved professionals in administrative careers, with normal or corrected-to normal vision, and naive to the scientific hypotheses underlying the experiments. Pre-screening interviews were conducted to make sure that none of the selected participants had any particular experience in knitting, eating with chopsticks, tool-mediated mechanical procedures, or surgery.

*Handedness, spatial ability, and study groups*

Participants' handedness was assessed using the Edinburgh inventory for handedness designed by [50] to confirm that they were all true right-handers. They were screened for spatial ability on the basis of the PTSOT (Perspective Taking Spatial Orientation Test) developed by Hegarty and Waller [49]. This test permits evaluating the ability of individuals to form three-dimensional mental representations of objects and their relative localization and orientation on the basis of merely topological (i.e. non-axonometric) visual data displayed two-dimensionally on a sheet of paper or a computer screen. All participants scored successful on 10 or more of the 12 items of the test, which corresponds to performances well above average. After pre-screening, they were divided at random into two groups of four men and four women each. Both groups performed the same tasks under the same conditions with the exception of that of the 2D monitor position, which varied between groups. The monitor was placed sideways for one group, and straight ahead for the other.

*Experimental platform: hardware and software*

The experimental platform is a combination of hardware and software components designed to test the effectiveness of varying visual environments for image-guided action in the real world (Fig 1). The main body of the device contains adjustable horizontal and vertical aluminium bars connected to a stable but adjustable wheel-driven sub-platform. The main body can be resized along two different axes in height and in width, and has two HD USB cameras (ELP, Fisheye Lens) fitted into the structure for monitoring the real-world action field from a stable vertical height, which was 60 cm here in this experiment. In this study here, a single 2D camera view was generated for the 2D monitor conditions through one of the two 120° fisheye lens cameras, both fully adjustable in 360°, connected to a small piece of PVC. For stereoscopic 3D viewing, views from the two cameras for left and right images were generated. The video input received from the cameras was processed by a DELL Precision T5810 model computer equipped with an Intel Xeon CPU E5-1620 with 16 Giga bytes memory (RAM) capacity at 16 bits and an NVidia GForce GTX980 graphics card. Experiments were programmed in Python 2.7 using the Open CV computer vision software library. The computer was connected to a high-resolution color monitor (EIZO LCD 'Color Edge CG275W') with an inbuilt color calibration device (colorimeter), which uses the Color Navigator 5.4.5 interface for Windows. The colors of objects visualized on the screen were matched to RGB color space, fully compatible with Photoshop 11 and similar software tools. The color coordinates for RGB triples were retrieved from a look-up table. The screen luminance values for calculating the object contrasts displayed for an experiment are given by the output of the EIZO auto-calibration procedure in *candela per square meter* ($cd/m^2$). All values were cross-checked with standard photometry using an external photometer (Cambridge Research Instruments) and interface software.

**Fig 1: Snapshot views of the experimental platform.** How individuals were seated, the two monitor positions (sideways and straight ahead), and examples of the 2D (top) and 2D fisheye (bottom) viewing conditions are shown. In the "natural" direct viewing condition (not shown here), the subject was positioned as in the Oculus head-mounted virtual 3D stereo viewing scenario (left).

*Objects in the real-world action field*

The Real-world Action Field (as of now referred to as the RAF) consisted of a classic square shaped (45cm x 45cm) light grey LEGO© board available worldwide in the toy sections of large department stores. Six square-shaped (4.5cm x 4.5cm) target areas were painted on the board at various locations in a medium grey tint (acrylic). In-between these target areas, small LEGO© pieces of varying shapes and heights were placed to add a certain level of complexity to both the visual configuration and the task and to reduce the likelihood of getting performance ceiling effects. The object that had to be placed on the target areas in a specific order was a small (3cm x 3cm x 3cm) cube made of very light plastic foam but resistant to deformation in all directions. Five sides of the cube were painted in the same medium grey tint (acrylic) as the target areas. One side, which was always pointing upwards in the task, was given an ultramarine blue tint (acrylic) to permit tracking object positions. A medium sized barbecue tong with straight ends was used for manipulating the object in the conditions 'with tool' (Fig 1). The tool-tips were given a matte fluorescent green tint (acrylic) to permit tool-tip tracking.

*Objects displayed on the 2D monitor*

The video input received by the computer from the HD USB camera generated the raw image data. These were adjusted to a viewing frame of 640 pixels (width) x 480 pixels (height) and processed to generate 2D visual displays in a viewing frame of 1280 pixels (width) x 960 pixels (height), the size of a single pixel on the screen being 0.32mm. Real-world data and visual display data were scaled psychophysically for each observer, i.e. the image size was adjusted for each subject to ensure that the visual display subjectively matched the scale of the RAF seen in the real world as closely as possible. A camera output matrix with image distortion coefficients using the Open CV image library in Python was used to correct the fisheye effects for the 2D undistorted viewing conditions of the experiment. The luminance ($\boldsymbol{L}$) of the light gray RAF visualized on the screen was 33.8 cd/m$^2$ and the luminance of the medium gray target areas was 15.4 cd/m$^2$, producing a target/background contrast (Weber contrast: $((\boldsymbol{L}_{\text{foreground}}-\boldsymbol{L}_{\text{background}})/\boldsymbol{L}_{\text{background}})$) of -0.54. The luminance of the blue (x=0.15, y=0.05, z=0.80 in CIE color space) object surface visualized on the screen was 3.44 cd/m$^2$, producing Weber contrasts of -0.90 with regard to the RAF, and -0.78 with regard to the target areas. The luminance (29.9 cd/m$^2$) of the green (x=0.20, y=0.70, z=0.10 in CIE color space) tool-tips produced Weber contrasts of -0.11 with regard to the RAF, and 0.94 with regard to the target areas. All luminance values for calculating the object contrasts visualized on the screen were obtained on the basis of standard photometry using an external photometer (Cambridge Research

Instruments) and interface software.

*Objects displayed in 3D through head-mounted OCULUS DK2*

The video input received by the computer from two HD USB cameras was fed into a computer vision software (written in Python 2.7 for Windows) which transforms the input data from the two cameras into a stereoscopic 3D image, displayed on the head-mounted screen of the OCULUS DK2 (www.oculus.com/dk2). Real-world data and visual display data were scaled psychophysically for each observer, i.e. the image size was adjusted for each subject to ensure that the visual display subjectively matched the scale of the RAF seen in the real world as closely as possible. In all the image-guided conditions (2D and stereoscopic 3D), image frames were displayed as quickly as possible at a frame rate of 30 Hz.

*Experimental procedure*

The experiments were run under conditions of free viewing, with illumination levels that can be assimilated to daylight conditions. The RAF was illuminated by two lamps (40Watt, 6500 K) which were constantly lit during the whole duration of an experiment. Participants were comfortably seated at a distance of approximately 75cm from the RAF in the direct viewing condition (Fig 1). For the group who performed the 2D image-guided conditions with the monitor placed sideways, there was a lateral angle of offset from the forearm motor axis of about 45° to the left (sub-optimal monitor position, see introduction), and the screen was about 75 cm away from their eyes (Fig 1). For the group who performed the 2D image-guided conditions with the monitor placed straight ahead of them, there was no lateral offset from the forearm motor axis, and the screen was about 150 cm away from their eyes (Fig 1). To compensate for the change in image size on the screen with the change in body-to-screen distance, the image on the screen was adjusted, ensuring that the perceived scale of the RAF displayed in an image subjectively matched the perceived scale of the RAF when viewed directly. Seats were adjusted individually in height at the beginning of a session to ensure that in both groups the image displayed on the monitor was slightly higher than the individual's eyes when looking straight at the screen, which is a near-optimal position given that the optimal monitor height is deemed to be one slightly lower than the eye-level (see introduction). All participants were given a printout of the targets-on-RAF configuration with white straight lines indicating the ideal object trajectory, and the ordered (red numbers) target positions the small blue cube object had to be placed on in a given trial set of the positioning task (Fig 2), always starting from zero, then going to one, to two, to three, to four, to five, and back to position zero. Participants were informed that they would have to position the cube with their dominant hand "as precisely as possible on the center of each target, as swiftly as possible, and in the right order, as indicated on the printout". They were also informed that they were going to be asked to perform this task under different conditions of object manipulation: with their bare right hand or using a tool, while viewing the RAF (and their own hand) directly in front of them, on a computer screen, or through the head-mounted Oculus device rendering a 3D image. All participants grasped the object with the thumb and the index of their right hand from the right-hand side in the bare-handed manipulation condition, and from the front with tongs held in their right hand when the tool was used. Before starting the first trial set, the participant could look at a printout of the RAF with the idel trajectory steps in the right order for as long as he/she wanted. When they felt confident that they remembered the target order well enough to do the task, the printout was taken away from them and the experiment was started. In the direct viewing condition, participants saw the RAF and what their hands were doing in top-down view through a glass window (Fig 1). In the other viewing conditions, the subjects had to look at a top-down 2D (fisheye and undistorted) or 3D image view of the RAF (Fig 1). Each participant was run in each of the different experimental conditions twice, in two separate successive sessions. A session always began with the condition of direct viewing, which is the easiest in the light of earlier findings [6]. Thereafter, the order of two 2D and 3viewing conditions (2D undistorted, 2D fisheye, 3D OCULUS) was counterbalanced, between sessions and between participants, to avoid order specific habituation effects. For the same reason, the order of the tool-use conditions (with and without tool) was also counterbalanced, between sessions and between participants. Given the four levels of the viewing factor ($V_4$: direct *vs* 2D undistorted *vs* 2D fisheye *vs* 3D OCULUS) combined with the two levels of the manipulation factor ($M_2$: no tool *vs* tool), the two levels of the gender factor ($G_2$: men, women) and the two levels of the session factor ($S_2$: first session *vs* second session), and with ten repeated trial sets per condition for four individuals of each gender in the two study groups, we have a Cartesian design plan with four principal design variables $V_4 \, x \, M_2 \, x \, G_2 \, x \, S_2$ and ten repeated trial sets in each condition and for each of the eight individuals from each of the two study groups with the two different monitor positions. The monitor position factor ($P_2$: sideways *vs* straight ahead) is our fifth principal design variable (*between-groups factor*).

**Fig 2. Screenshot view of the RAF.** The ideal object trajectory, from position zero to positions one, two, three, four, five, and then back to zero, is indicated by the white line here. Participants had to pick and place a small foam cube with blue top on the centers of the grey target areas in the order shown here, as precisely as possible and as swiftly as possible.

*Data generation*

Data from fully completed trial sets only were recorded. A fully complete trial set consists of a set of positioning operations starting from zero, then going to one, to two, to three, to four, to five, and back to position zero without dropping the object accidentally, and without errors in the positioning order. Whenever such occurred (this happened only incidentally, mostly at the beginning of the experiment), the trial set was aborted immediately, and the participant started from scratch in that specific condition. As stated above, ten fully completed trial sets were recorded for each combination of factor levels. For each of such ten trial sets, the computer program generated data relative to the dependent variables 'time' and 'precision'. The computer vision software, written specifically for this experiment in Python, took care of aligning the video image data with the real-world data and counted the task execution time of each individual trial. This execution time corresponds to the CPU time (in milliseconds) from the moment the blue cube object was picked up by the participant to the time it was put back to position zero again. The frame rate of 2D images was between 25-30 Hz, with an error margin of less than 40 milliseconds for any of the time estimates. Each frame was processed individually for data collection. For the precision estimates, the computer program counted the cumulated number of blue object pixels at positions "off" the 3cm x 3cm central area of each of the five 4.5cm x 4.5cm target areas whenever the object was positioned on a target. The standard errors of these positional estimates, determined in a calibration procedure, were below 10 pixels. "Off-center" pixels were not counted for object positions on the square labeled 'zero' (the departure and arrival square). Individual time and precision data were written to an excel file by the computer program and stored in a directory for subsequent analysis.

# Results and discussion

Means and standard errors for each of the two dependent variables ('time' and 'precision') were computed for a first scrutiny, and then the raw data were submitted to analysis of variance (ANOVA).

In a first step, the data from the two study groups with the different 2D monitor positions were grouped together to assess the effects of the inter-group factor $P_2$ (monitor position). A 5-way analysis of variance (ANOVA) was run in MATLAB (7.14) on raw data for ‘time’ and ‘precision’. This analysis took into account only the two 2D conditions of the viewing factor (2D undistorted *vs* 2D fisheye) in combination with the two levels of the monitor position factor (straight ahead *vs* sideways), the two levels of the manipulation factor (tool *vs* no tool), the two levels of the gender factor (men *vs* women), and the two levels of the session factor (first session *vs* second session). Given ten repeated trial sets per condition with four men and four women in each of the two study groups, we have the following five-factor analysis: $V_2$ x $P_2$ x $M_2$ x $G_2$ x $S_2$ combined with 10 repeated sets for the four individuals per gender and a total of 1280 raw data for ‘time’ and for ‘precision’. Table 1 summarizes the results of this first analysis, showing means and standard errors for the different experimental conditions (effect sizes), and *F* and *p* values signaling the statistical significance of the effect of each principal design variable (factor) on the dependent variables ‘time’ and ‘precision’.

**Table 1. Means, standard errors, *F* and *p* values from the 5-Way ANOVA.**

| **MAIN FACTOR EFFECT F and *p* values** | FACTOR LEVELS | AVERAGE ***TIME*** OF TASK EXECUATION (SECONDS) | |
|---|---|---|---|
| | | MEAN | SEM |
| **Viewing** F (1,1279)=2.36; *NS* | 2D | 11.98 | 0.11 |
| | 2D fisheye | 12.20 | 0.14 |

| **Monitor Position** F(1,1279)=318; *p* <.001 | Ahead | 10.60 | 0.10 |
|---|---|---|---|
| | Sideways | 13.60 | 0.15 |
| **Manipulation** F(1,1279)=77.95; *p* < .001 | No Tool | 11.30 | 0.10 |
| | Tool | 10.60 | 0.15 |
| **Session** F(1,1279)=218.15; *p* < .001 | Session 1 | 13.30 | 0.12 |
| | Session 2 | 10.80 | 0.10 |
| **Gender** F(1, 1279)=11.73; *P* < .01 | Male | 11.80 | 0.12 |
| | Female | 12.40 | 0.10 |

| **MAIN FACTOR EFFECT F and *p* values** | FACTOR LEVELS | AVERAGE ***PRECISION*** (NUMBER OF PIXELS "OFF" TARGET CENTER) | |
|---|---|---|---|
| | | MEAN | SEM |
| **Viewing** F (1,1279)=22.97; *p* < .001 | 2D | 1120 | 14 |
| | 2D fisheye | 1010 | 16 |
| **Monitor Position** F(1,1279)=23.82; *p* <.001 | Ahead | 1121 | 15 |
| | Sideways | 1009 | 17 |
| **Manipulation** F(1,1279)=1.46; *NS* | No Tool | 1079 | 16 |
| | Tool | 1051 | 16 |
| **Session** F(1,1279)=0.63; *NS* | Session 1 | 1074 | 17 |
| | Session 2 | 1056 | 15 |
| **Gender** F(1, 1279)=0.73; *NS* | Male | 1055 | 18 |
| | Female | 1075 | 16 |

5-way ANOVA on data from both study groups put together was computed for comparing between the 2D monitor viewing conditions for 'time' (top) and 'precision' (bottom).

The results for 'time' show no effect of 2D undistorted *vs* 2D fisheye, but significant effects of monitor position, manipulation, session, and gender. Subjects were significantly faster in the group where the monitor was placed straight ahead of them. They were significantly faster when no tool was used to perform the task. Times are significantly shorter in the second session compared with the first (training effect). Men executed the tasks significantly faster than the women. Results for ‘precision’ show a significant effect of viewing where 2D fisheye viewing yields a significantly better precision score than 2D undistorted viewing. Subjects were significantly more precise in the group where the monitor was positioned sideways. Neither the manipulation mode, nor the session factor (training), nor gender had any significant effect on ‘precision’ in this analysis. There were no significant two-way interactions between factors.

In a second step, the data from each study group were analyzed separately. Descriptive analyses were performed first, and boxplots showing the data distributions around the medians in the four different viewing conditions, for each study group separately, were generated (Fig 3). Outliers in the data were indeed rare and given the large amount of data collected for each condition, correcting these few by replacing them by averages would not have changed the statistical analyses. The raw data for each group were therefore submitted to ANOVA as shown here.

**Fig 3. Data distributions around the medians in the four different viewing conditions.** Data for ‘time’ (top) and ‘precision’ (bottom) from the group with the 2D monitor positioned straight ahead are shown on the left and data from the group with the monitor positioned sideways are shown on the right. Note that in the stereoscopic 3D condition (v4), the viewing monitor is head-mounted and moves along naturally with the head of the subject.

4-way ANOVA was performed on raw data for ‘time’ and ‘precision’ from each of the two study groups independently. These analyses took into account all four conditions of the viewing factor (direct *vs* 2D *vs* 2D fisheye *vs* 3D head-mounted) for each study group in combination with the two levels of the two levels of the manipulation factor (tool *vs* no tool), the two levels of the gender factor (men *vs* women), and the two levels of the session factor (first session *vs* second session). Given ten repeated trial sets per condition with four men and four women in each of the two study groups, we have the following four-factor analysis: $V_4$ x $M_2$ x $G_2$ x $S_2$ combined with 10 repeated sets for the four individuals per gender and a total of 1280 raw data for 'time' and for 'precision'. Tables 2 and 3 summarize the results of these analyses, showing means and standard errors for the different experimental conditions (effect sizes), and *F* and *p* values signaling the statistical significance of the effect of each principal design variable (factor) on the dependent variables 'time' and 'precision'.

**Table 2. Means, standard errors, and *F* and *p* values from the 4-way ANOVA**.

| **MAIN FACTOR EFFECT F and *p* values** | FACTOR LEVELS | AVERAGE ***TIME*** OF TASK EXECUATION (SECONDS) | |
|---|---|---|---|
| | | MEAN | SEM |
| **Viewing** F (3,1279)=301.62; *p* < .001 | Direct | 6.03 | 0.09 |
| | 2D | 10.61 | 0.14 |
| | Oculus 3D | 11.38 | 0.17 |
| | 2D fisheye | 10.57 | 0.16 |
| **Manipulation** F(1,1279)=81.51; *p* < .001 | No Tool | 9.03 | 0.12 |
| | Tool | 10.26 | 0.14 |
| **Session** F(1,1279)=72.29; *p* < .001 | Session 1 | 10.22 | 0.14 |
| | Session 2 | 9.07 | 0.12 |
| **Gender** F(1, 1279)=1.87; *NS* | Male | 9.57 | 0.13 |
| | Female | 9.73 | 0.14 |

| **MAIN FACTOR EFFECT F and *p* values** | FACTOR LEVELS | AVERAGE ***PRECISION*** (NUMBER OF PIXELS "OFF" TARGET CENTER) | |
|---|---|---|---|
| | | MEAN | SEM |
| **Viewing** F (3,1279)=200.86; *p* | Direct | 719 | 21 |
| | 2D | 1140 | 23 |

| < .001 | Oculus 3D | 1645 | 36 |
|---|---|---|---|
| | 2D fisheye | 1100 | 28 |
| **Manipulation** F(1,1279)=37.05; *p* < .001 | No Tool | 1070 | 22 |
| | Tool | 1233 | 25 |
| **Session** F(1,1279)=2.31; *NS* | Session 1 | 1131 | 24 |
| | Session 2 | 1172 | 23 |
| **Gender** F(1, 1279)=0.50; *NS* | Male | 1141 | 22 |
| | Female | 1162 | 25 |

Results for 'time' (top) and 'precision' (bottom) of the study group with the 2D monitor positioned straight ahead. In the Oculus 3D condition, the monitor was head-mounted.

**Table 3. Means, standard errors, and *F* and *p* values from the 4-way ANOVA**.

| **MAIN FACTOR EFFECT F and *p* values** | FACTOR LEVELS | AVERAGE ***TIME*** OF TASK EXECUATION (SECONDS) | |
|---|---|---|---|
| | | MEAN | SEM |
| **Viewing** F (3,1279)=321.90; *p* < .001 | Direct | 7.36 | 0.11 |
| | 2D | 13.32 | 0.19 |
| | Oculus 3D | 12.95 | 0.19 |
| | 2D fisheye | 13.87 | 0.24 |
| **Manipulation** F(1,1279)=94.49; *p* < .001 | No Tool | 11.06 | 0.16 |
| | Tool | 12.70 | 0.17 |
| **Session** F(1,1279)=195.42; *p* < .001 | Session 1 | 13.06 | 0.18 |
| | Session 2 | 10.69 | 0.14 |
| **Gender** F(1, 1279)=34.72; *NS* | Male | 11.38 | 0.16 |
| | Female | 12.38 | 0.18 |

| **MAIN FACTOR EFFECT F and *p* values** | FACTOR LEVELS | AVERAGE ***PRECISION*** (NUMBER OF PIXELS “OFF” TARGET CENTER) | |
|---|---|---|---|
| | | MEAN | SEM |
| **Viewing** F (3,1279)=166.06; *p* < .001 | Direct | 537 | 16 |
| | 2D | 1099 | 20 |
| | Oculus 3D | 1115 | 28 |
| | 2D fisheye | 920 | 21 |
| **Manipulation** F(1,1279)=4.19; | No Tool | 939 | 18 |
| | Tool | 896 | 18 |

| *p* < .05 | | | |
|---|---|---|---|
| **Session**<br>F(1,1279)=0.63;<br>*NS* | Session 1 | 927 | 18 |
| | Session 2 | 908 | 17 |
| **Gender**<br>F(1, 1279)=14.77;<br>*P* < .001 | Male | 958 | 19 |
| | Female | 878 | 16 |

Results for 'time' (top) and 'precision' (bottom) of the study group with the 2D monitor positioned sideways. In the Oculus 3D condition, the monitor was head-mounted.

Results for ‘time’ from the group with the monitor positioned straight ahead (Table 2) and from the group with the monitor positioned sideways (Table 3) show quite clearly and consistently that subjects in both groups performed significantly faster in the direct viewing condition, and took significantly more time in all the four image viewing conditions, which produced roughly equivalent data for ‘time’ in each of the two groups. The sideways group (Table 3) took on average two seconds longer than the *straight ahead* group (Table 2) in all the experimental conditions. The manipulation factor also affected both study groups in the same way, as subjects from both groups performed significantly faster when they did not have to use a tool. Subjects from both study groups were significantly faster in the second session compared with the first (i.e. we have a training effect on ‘time’). There is no difference between the task execution times of men and women in the group with the monitor positioned straight ahead. In the *sideways* group, the men performed significantly faster than the women. Results for ‘precision’ from the group with the monitor positioned straight ahead (Table 2) and from the group with the monitor positioned sideways (Table 3) show quite clearly and consistently that subjects in both groups were significantly more precise in the direct viewing condition than in any of the image viewing conditions, which produced roughly equivalent data for ‘precision’ in each of the two groups. The *sideways* group (Table 3) was more precise than the *straight ahead* group (Table 2) in all the experimental conditions. The manipulation factor also affected both study groups in the same way, as subjects from both groups were significantly more precise when they did not have to use a tool. Subjects from neither study group were more precise in the second session compared with the first (i.e. we have no training effect on 'precision'). There is no difference between the precision scores of men and women in the group with the monitor positioned straight ahead. In the *sideways* group, the women were significantly more precise than the men. Interactions are not shown in the Tables. We found significant interactions between the viewing and the manipulations factors in each of the two study groups (Fig 4).

**Fig 4. Interactions.** Task execution times (top) and pixel-based precision parameters (bottom) are shown as a function of the four different viewing conditions and the two manipulation conditions, for the *straight ahead* group (left) and the *sideways* group (right).

In the *straight ahead* group, there was no significant interaction between viewing and tool-use in their effects on 'time' ($F(3,1279)=2.06$; NS). We found such an interaction in the *sideways* group ($F(3,1279)=4.17$; $p<.01$), independent of the change in 2D monitor position (Fig 4) The interaction only involves the head-mounted 3D viewing condition, where the tool-use has a more detrimental effect on times than in any of the other viewing conditions. In both study groups, we found significant interactions between viewing and tool-use in their effects on 'precision' ($F(3,1279)=7.30$; $p<.001$ in the *straight ahead* group and $F(3,1279)=5.15$; $p<.01$ in the *sideways* group), involving the head-mounted 3D and the 2D fisheye viewing conditions (Fig 4).

The results show that, compared with the direct viewing condition, the three image viewing conditions had significantly detrimental effects on the time and the precision with which the participants placed the small cube object on the target centers in the specific order. The negative effects of 2D image views compared with "natural" direct action viewing were predicted on the basis of earlier findings from the seminal studies by Hubber and colleagues [5] and Gallagher and colleagues [6], which made a strong impact by showing that 2D image-guided performance is never as good as performance guided by natural human vision, for reasons beyond loss of binocular disparity information available in natural viewing. The absence of a superiority effect of head-mounted 3D viewing compared with 2D viewing from different monitor positions in our data is consistent with previous findings by some authors [19-21], and in seeming contradiction with data from studies published by others showing such a superiority effect [12-17]. The major implications of these findings will be discussed in detail in the following paragraphs.

*2D fisheye vs undistorted 2D*

Although the 2D fisheye viewing condition would have been expected to affect performances more negatively than undistorted 2D screen viewing, the opposite was observed. Given the task instruction to place the cube as precisely as possible on the target centers, the 2D fisheye version of the RAF may have generated a task-specific facilitation effect on precision. In fact, in the top-down 2D fisheye view, the targets appear dome-like rather than flat, as in top-down undistorted 2D viewing, which makes the target centers perceptually more distinguishable from the image background.

*Monitor position*

The between-groups factor monitor position affected performances significantly, but in opposite directions for task execution times and the pixel-based precision score: while subjects performed significantly faster in the two 2D viewing conditions in the group with the monitor positioned straight ahead, they were also significantly less precise in that group. This is an important finding because it suggests that subjective comfort factors need to be considered in tight relation to individual goal-setting strategies [7, 8, 10]. Subjects in the straight ahead group experienced less strain on the neck during task execution, as previously reported [25], and therefore felt more comfortable and fully disposed to go as fast as they could, while the subjects in the other group felt less comfortable [26, 27] and therefore paid more attention to the precision of their manoeuvers. Trade-off effects between speed and precision of task execution are an important aspect of the performances of novices and well-known to reflect individual strategy variations [51-57]. These strategy variations are difficult to predict in complex tasks because they do not depend on any single parameter, or clearly identified factor combination. They result from a multitude of internal and external constraints. State-of-the-art research in the neurosciences of goal-related strategies and decision making suggests that they are top-down controlled by the temporal lobes of the human brain [58, 59].

*Stereoscopic 3D vs 2D*

Stereoscopic 3D viewing through the head-mounted device did not represent a performance advantage compared with the 2D image viewing conditions in this study here. In some of the earlier studies, authors concluded that novice and expert users with normal capacity for spatial perception can work faster and safer under 3-D vision, especially in complicated surgical tasks [14, 18, 24]. Several explanations may account for the difference between these and our results here.

First, most of the previous studies were run on surgeons with different levels of expertise, from so-called novices to so-called experts. It is difficult to render novice groups from a population consisting of professional surgeons homogenous with respect to eye-hand coordination expertise. All surgeons are experts in this regard, yet, they are more or less proficient at different specific tasks. This variability may not be easy to track down. For this reason, our experiment here was run on complete novices, all scoring high in spatial ability, without any surgical experience at all.

Second, high resolution 2D/3D surgical camera systems, as the one used in one of the most recent studies [18], not only control for viewing angle and distance like our camera display here, but these cameras also move along with the tools during task execution. Our cameras had fixed locations. When the cameras are moving along with the tool, the movements represented visually are not aligned with the surgeon's real arm and hand movements. Thus, when such a system is switched into 3D mode, the stereoscopic information conveyed could help overcome this problem, which would explain why task execution is easier, especially for the less trained surgeons, compared with the 2D mode [18].

Third, in our display here, the tool-tips and a critical part of the manipulated object (the top) were selectively coloured for tracking. These colours may have provided particularly powerful visual cues for task execution in 2D [60-62], cancelling the major advantage of stereoscopic viewing. Studies in image-guided neurosurgery [60, 61] have previously shown that adding colour to specific locations in 2D images produces strong and self-sufficient cues to visual depth for interventional guidance, especially in novices, potentially making 3D viewing unnecessary.

Finally, the absence of a 3D superiority effect here in our study may be partly be due to the complex interactions between viewing and manipulation modalities, i.e. the tool-use factor, affecting subjectively extended near-body space [43, 45]. Absolute beginners from possibly heterogeneous general training backgrounds have to learn to adjust to extended near-body space when using a tool, especially when confronted with different viewing modalities. These complex processes of adjustment have not yet been studied in the context of image-guided eye-hand coordination, and more research oriented in that direction is needed.

*Interactions between viewing and tool-use factors*

The performances of both the men and the women were significantly impaired when they had to use a tool to perform the positioning task compared with the conditions where they used their bare hand. Tool-specific motor requirements [46, 63], such as having to grab and hold the handle of the tool, or having to adjust one's hand movements to the shape and the size of the tool, would readily account for this effect. However, given the significant interaction of this effect with the effects of the different viewing conditions found here, we clearly need more knowledge about how different viewing modalities affect so-called near-body space. The latter is defined as the space around one's own body within arm's reach and its perceived extent affects performance by drawing attention to regions of space that are not paid attention to when the same task is performed with the hands directly [44]. Body space extension through the tool explains why it is easier to position an object with a tool in far-away space, but we do not know how this space scales in different 2D and 3D viewing conditions.

*Gender effect or inter-individual strategy differences?*

The gender effect showing that men performed significantly faster than the women has to be interpreted with much caution. First, other studies have shown effects in the opposite direction, reporting faster performance in women compared with men [64]. Second, temporal performance scores must not be considered without taking into account the precision scores, for reasons already pointed out here above and explained in terms of individually specific goal-related speed-accuracy trade-offs. These depend on the type of task, and on other, physiological and psychological, factors which need to be identified. In this study here, it is shown that the men were significantly faster but, at the same time, also significantly less precise than the women in the sideways group. This apparent gender effect is absent in the straight ahead group, but cannot be explained away by the mere difference in monitor position. Monitor position affects subjective comfort levels [21, 25, 65], and subjective comfort levels affect individual goal-setting, which involves criteria for timing and precision strategies [66, 67]. More research is clearly needed to understand these complex processes.

# Conclusions

In consistency with earlier findings, image-guidance significantly slows down, and significantly reduces the precision of, goal-directed manual operations of novices, all non-surgeons scoring high in spatial ability. In seeming contradiction with some of the results reported previously, we found no superiority of stereoscopic 3D image viewing (head-mounted device OCULUS DK2) compared with 2D viewing. This result may be explained by a combination of effects relative to the study population, the camera systems, and the specific colour cues made available for tracking the tooltips and the object manipulated here in this study, which provided powerful cues to visual depth in the 2D images. The complex interactions between viewing, tool-use, and individual strategy factors [68, 69, 70], expressed here in terms of an apparent gender effect, open new and important perspectives for further research on novices in image-guided eye-hand coordination.